\documentclass[aps,pra,twocolumn,superscriptaddress]{revtex4-2}

\usepackage{graphicx} 
\usepackage{float}
\usepackage{comment}
\usepackage{siunitx}
\usepackage{amsmath}
\usepackage{amssymb}
\usepackage{xcolor}
\usepackage{capt-of} 

\usepackage[
  colorlinks=true,
  linkcolor=blue,
  citecolor=blue,
  urlcolor=blue,
  pdfauthor={M. Ebrahimi et al.},
  pdftitle={Coherent Perfect Absorption: Zero Reflection Without Linewidth Suppression}
]{hyperref}
\usepackage[capitalise,nameinlink,noabbrev]{cleveref}
\crefname{figure}{Fig.}{Figs.}
\Crefname{figure}{Figure}{Figures}
\crefname{equation}{Eq.}{Eqs.}
\Crefname{equation}{Equation}{Equations}
\crefname{section}{Sec.}{Secs.}
\Crefname{section}{Section}{Sections}

\begin{document}

\title{Coherent Perfect Absorption: Zero Reflection Without Linewidth Suppression}

\author{M.~Ebrahimi}
\affiliation{Department of Physics, University of Alberta, Edmonton, Alberta T6G~2E9, Canada}

\author{Y.~Huang}
\affiliation{Department of Physics, University of Alberta, Edmonton, Alberta T6G~2E9, Canada}

\author{A.~Rashedi}
\affiliation{Department of Physics, University of Alberta, Edmonton, Alberta T6G~2E9, Canada}

\author{J.~P.~Davis}
\email[Corresponding author: ]{jdavis@ualberta.ca}
\affiliation{Department of Physics, University of Alberta, Edmonton, Alberta T6G~2E9, Canada}

\begin{abstract}
Motivated by recent claims, we revisit how coherent perfect absorption (CPA) influences cavity and polaritonic linewidths. Using standard input–output theory and measurements on single-port bare microwave cavities and cavity–magnon hybrids, we find that CPA drives the on-resonance reflection to zero while the spectral width remains set by the total decay rate. Apparent narrowing observed near CPA  is found to be a visual artifact that does not remain upon quantitative analysis. Extending the analysis to cavity magnomechanics, we show that logarithmic plots can exhibit apparent polaromechanical “normal-mode splitting,” whereas linear-scale spectra display no true splitting. These results clarify when CPA modifies amplitudes versus spectral poles, offer practical guidance for data presentation, and indicate that CPA alone is not a route to linewidth suppression or polaromechanical mode splitting in the linear, weak-probe regime. 
\end{abstract}

\maketitle


\section{Introduction}

Hybrid quantum systems provide a versatile platform for exploring quantum phenomena and advancing quantum information processing \cite{Xiang2013, Rogers2014, Kurizki2015}. By coherently interfacing disparate degrees of freedom—optical or microwave photons, superconducting circuits, and collective spin or mechanical excitations—these systems enable controlled energy exchange and access to nonclassical states. Reaching the strong-coupling regime, where the coherent interaction rate exceeds the relevant dissipation rates, is a central objective \cite{Zhang2014, Tabuchi2014, Potts2020, Grblacher2009, Teufel2011, Rossi2018}.

In many platforms, the intrinsic interaction is weak. For example, in cavity optomechanics the radiation-pressure coupling typically requires strong coherent driving to realize an effective, linearized interaction, at the cost of added noise and possible parametric instabilities \cite{Kippenberg2008, Ludwig2008, Kippenberg2009, Qian2012, Aspelmeyer2014}. Magnon-based hybrids—cavity magnonics and cavity magnomechanics—have recently emerged as compelling alternatives \cite{Huebl2013, Zhang2014, Tabuchi2014, Goryachev2014, Zhang2017, Potts2020, ZareRameshti2022, Zhang2016, Potts2021}. In these systems, microwave photons couple strongly to magnons in ferrimagnets such as yttrium iron garnet (YIG), forming cavity–magnon polaritons \cite{Huebl2013, Zhang2014, Tabuchi2014, Goryachev2014, Zhang2017, Potts2020, ZareRameshti2022}. Via magnetostriction, polaritons can further couple to mechanical vibrations \cite{Zhang2016, Potts2021}, opening a path to polaromechanics and, ultimately, to quantum control of phonons.

Beyond driven-enhanced coupling, dissipation/reservior engineering has been used to tailor systems dynamics—e.g., to create entangled steady states or enable quantum-limited amplification and nonreciprocity—\cite{Woolley2014, Metelmann2014, Metelmann2015, Fedorov2019}, and to improve cooperativity by reducing intrinsic or effective linewidths \cite{ Rossi2017, Rossi2018, Fedorov2019, Shen2025CavityMagnon}.
In particular, Ref.~\cite{Shen2025CavityMagnon} argued that tuning the external coupling of a cavity-magnonics to realize coherent perfect adsorption (CPA), which has been used advantageously in optics \cite{Chong2010, wan2011, Noh2012, Wang2021}, reduces the polariton linewidth and, in turn, enables polaromechanical normal-mode splitting.

Here, we revisit how CPA affects spectral widths in the linear, weak-probe regime. We begin with a single-port bare microwave cavity in which the external coupling (e.g., pin depth) sets the external dissipation rate. Using standard input–output theory together with measurements, we find that CPA coincides with critical coupling at resonance: destructive interference drives the on-resonance reflection to zero, while the spectral width remains fixed by the total decay rate $\kappa_a=\kappa_{\mathrm{int}}+\kappa_{\mathrm{ext}}$, where $\kappa_\mathrm{int}$ and $\kappa_\mathrm{ext}$ indicate the internal and external dissipation rates, respectively. As CPA drives the minimum of the reflection spectrum toward zero, logarithmic plots make the dip appear visually sharper and can give the impression of linewidth narrowing. The same spectra plotted in linear units show that the full width at half maximum (FWHM) is set by $\kappa_a$ and, for fixed $\kappa_{\mathrm{int}}$, is largest in the overcoupled regime and smallest in the undercoupled regime, confirming that the cavity linewidth is not reduced at CPA. Our contribution is to show, theoretically and experimentally, that CPA modifies on-resonance amplitudes but not the spectral poles, and to provide practical guidance for plotting and linewidth extraction.

We then apply the same analysis to a cavity-magnonics system. Again, although logarithmic plots can suggest linewidth narrowing at CPA, linear-scale spectra together with input–output theory confirm that CPA does not narrow the polariton linewidth; the damping remains set by the dissipation rate of the hybrid mode.

Finally, in the context of cavity magnomechanics, we show that logarithmic plots can also suggest an apparent polaromechanical “normal-mode splitting” under CPA, whereas linear-scale spectra display no true splitting. In these systems, CPA modifies amplitudes at resonance without shifting the spectral poles that determine the linewidth; hence no linewidth narrowing or splitting attributable to CPA is observed.


\begin{figure}[tbp]
    \centering
    \includegraphics[width=0.7\linewidth]{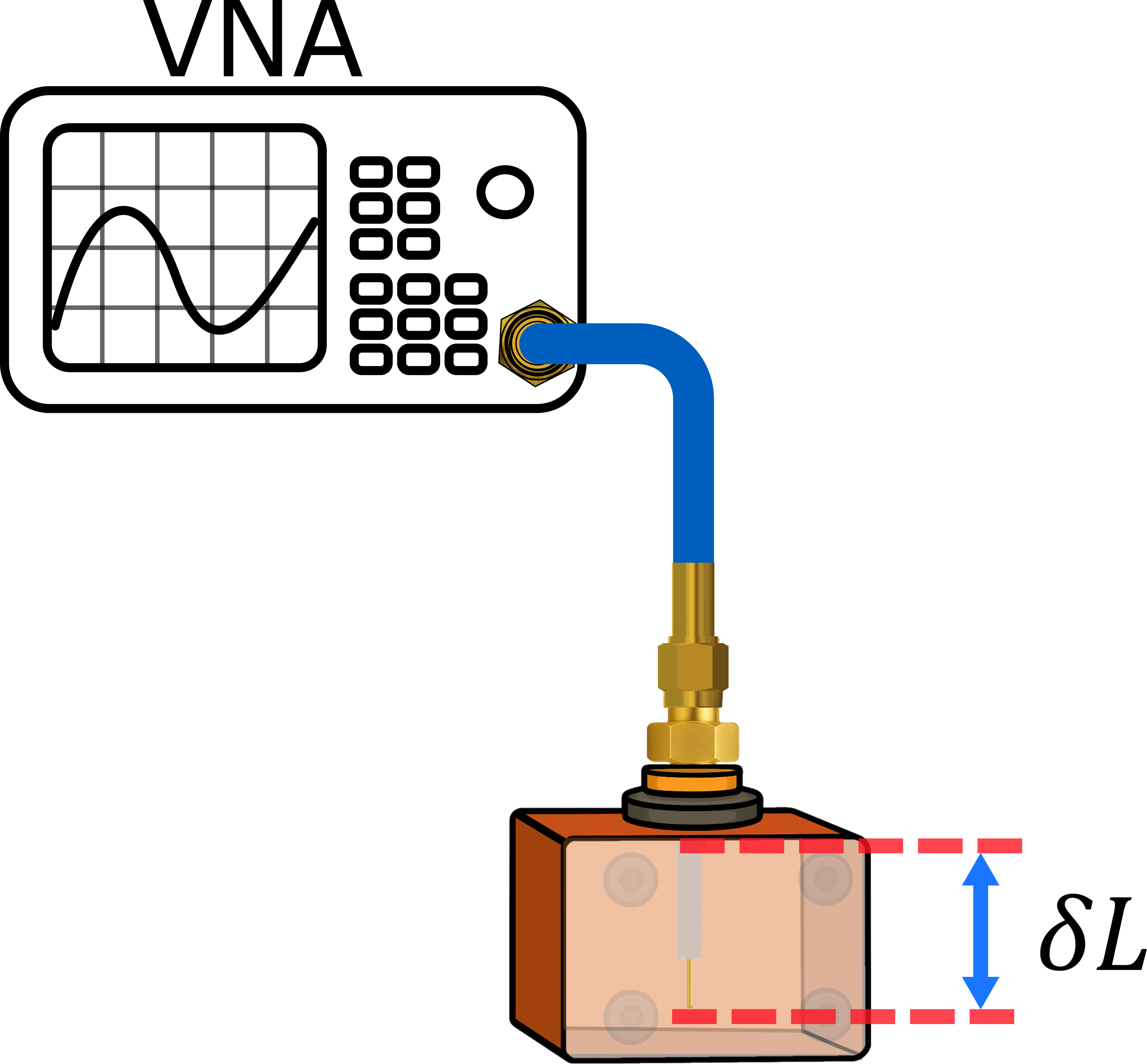}
    \caption{Schematic of the single-port reflection measurement of a bare cavity. The external coupling is tuned by adjusting the adapter-pin insertion depth.}
    \label{fig1}
\end{figure}


\section{Reflection spectra and CPA in a single-port bare microwave cavity}
\label{sec:bare-CPA}

\subsection{Experimental data}

\begin{figure}[tbp]
    \centering
    \includegraphics[width=\linewidth]{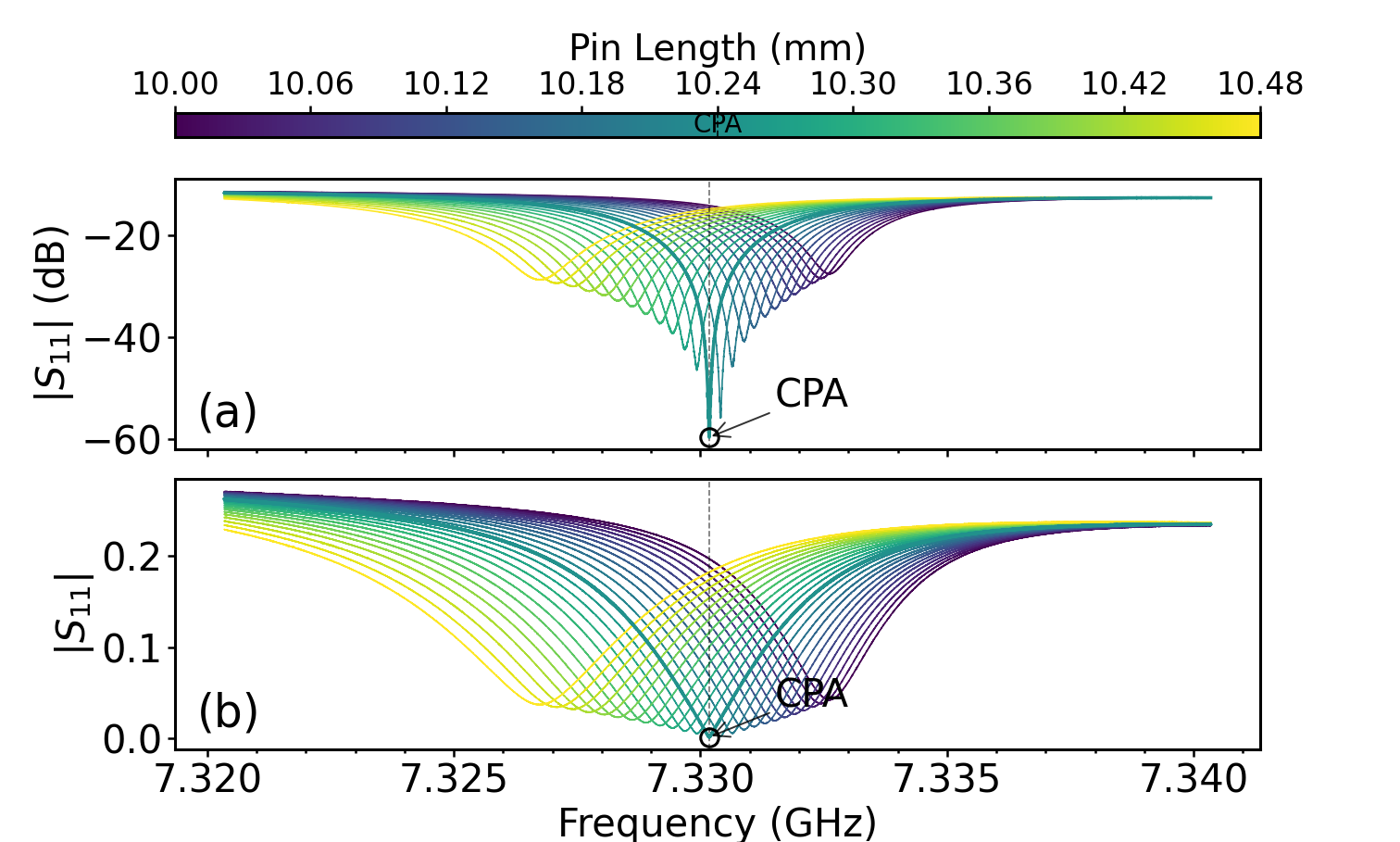}
    \caption{Reflection spectra of a single-port bare microwave cavity for several pin insertion length  $\delta l$
    (equivalently, external dissipation rate $\kappa_{\mathrm{ext}}$): (a) magnitude in logarithmic and (b) magnitude in linear units.
    The arrow marks coherent perfect absorption (CPA), realized at critical coupling
    $\kappa_{\mathrm{ext}}=\kappa_{\mathrm{int}}$, where the on-resonance dip ideally vanishes.
    The linewidth scale is set by the total decay rate
    $\kappa_a=\kappa_{\mathrm{int}}+\kappa_{\mathrm{ext}}$.}
    \label{fig2}
\end{figure}

We consider a single-port bare microwave cavity with resonance frequency $\omega_a/2\pi \approx \SI{7.33}{\giga\hertz}$. The input–output coupling is tuned by adjusting the insertion depth of the SMA adapter’s center pin [Fig.~\ref{fig1}], which sets the external dissipation rate $\kappa_{\mathrm{ext}}$. Representative reflection spectra $|S_{11}|$ measured with a vector network analyzer (VNA) are shown in Fig.~\ref{fig2} on both log and linear scales. As the coupling is varied, the on-resonance reflection can be driven close to zero by destructive interference between the incident wave and the field leaking from the cavity—this is identical to the concept of CPA in optical systems \cite{Chong2010, wan2011, Noh2012, Wang2021}. When displayed logarithmically, the deep dip at CPA appears visually “sharper,” which can suggest a narrowed linewidth; plotting the same data in linear units makes clear that the full width at half maximum (FWHM) does not shrink at CPA.

\subsection{Theoretical model}

Now we theoretically derive the reflection spectrum of a single-port bare microwave cavity. We then analyze the conditions for zero reflection, $S_{11}(\omega_p)=0$, and compare the theoretical results with the experimental data extracted from Fig.~\ref{fig2}.
With a weak probe at frequency $\omega_p$ and complex amplitude $\varepsilon_p$, the driven cavity is described by
\begin{equation}
\frac{H}{\hbar}=\omega_a \hat a^\dagger \hat a
+i\sqrt{\kappa_{\mathrm{ext}}}\,\varepsilon_p\!\left(\hat a^\dagger e^{-i\omega_p t}-\hat a\,e^{i\omega_p t}\right).
\end{equation}
The corresponding Langevin equation for the intracavity field is
\begin{equation}
\dot{\hat a}=-(i\omega_a+\kappa_a/2)\,\hat a
+\sqrt{\kappa_{\mathrm{ext}}}\,\varepsilon_p e^{-i\omega_p t},
\end{equation}
with $\kappa_a\equiv \kappa_{\mathrm{int}}+\kappa_{\mathrm{ext}}$. Using the Fourier transform, the above equation is obtained as
\begin{equation}
    -i\omega_p a(\omega_p) = -(i\omega_a+\kappa_a/2)a(\omega_p)+\sqrt{\kappa_{ext}}\varepsilon_p.
    \label{eq:a-in-freq}
\end{equation}
The steady-state intracavity amplitude is
\begin{equation}
a(\omega_p)=\frac{\sqrt{\kappa_{\mathrm{ext}}}\,\varepsilon_p}{\kappa_a/2+i(\omega_a-\omega_p)}.
\end{equation}
In a one-port reflection geometry, the input-output theory \cite{Clerk2010}, 

\begin{equation}
    a_{\mathrm{out}} = \varepsilon_p-\sqrt{\kappa_{\mathrm{ext}}}a
    \label{input-output}
\end{equation}

leads to the following reflection spectrum
\begin{equation}
S_{11}(\omega_p)\equiv \frac{a_{\mathrm{out}}}{\varepsilon_p}
=\frac{\kappa_{\mathrm{ext}}-\kappa_{\mathrm{int}}-2i(\omega_a-\omega_p)}
{\kappa_{\mathrm{ext}}+\kappa_{\mathrm{int}}+2i(\omega_a-\omega_p)},
\label{eq:S11-amp}
\end{equation}
and
\begin{equation}
|S_{11}(\omega_p)|^2
=\frac{4(\omega_a-\omega_p)^2+(\kappa_{\mathrm{ext}}-\kappa_{\mathrm{int}})^2}
{4(\omega_a-\omega_p)^2+(\kappa_{\mathrm{ext}}+\kappa_{\mathrm{int}})^2}.
\label{eq:S11-power}
\end{equation}

Equations~\eqref{eq:S11-amp}–\eqref{eq:S11-power} make two points explicit:
(i) CPA ($S_{11}(\omega_p)=0$ / zero reflection) occurs only at resonance,
\begin{equation}
   \omega_p=\omega_a,
   \label{resonance-bare}
\end{equation}
and at critical coupling
\begin{equation}
    \kappa_{\mathrm{ext}}=\kappa_{\mathrm{int}},
    \label{critical-bare}
\end{equation}
(ii) the spectral width is set by the pole of the response,

\begin{equation}
    \tilde{\omega}_a=\omega_a-i\kappa_a/2,
    \label{pole-barecavity}
\end{equation}

hence by the total decay rate
$\kappa_a=\kappa_{\mathrm{int}}+\kappa_{\mathrm{ext}}$, independent of the numerator. Therefore,
achieving CPA does not reduce the cavity linewidth. For fixed $\kappa_{\mathrm{int}}$, the FWHM
(in angular frequency) equals $\kappa_a$ and is smallest in the undercoupled regime
$\kappa_{\mathrm{ext}}<\kappa_{\mathrm{int}}$ and largest in the overcoupled regime
$\kappa_{\mathrm{ext}}>\kappa_{\mathrm{int}}$.

\begin{figure}[t]
    \centering
    \includegraphics[width=\linewidth]{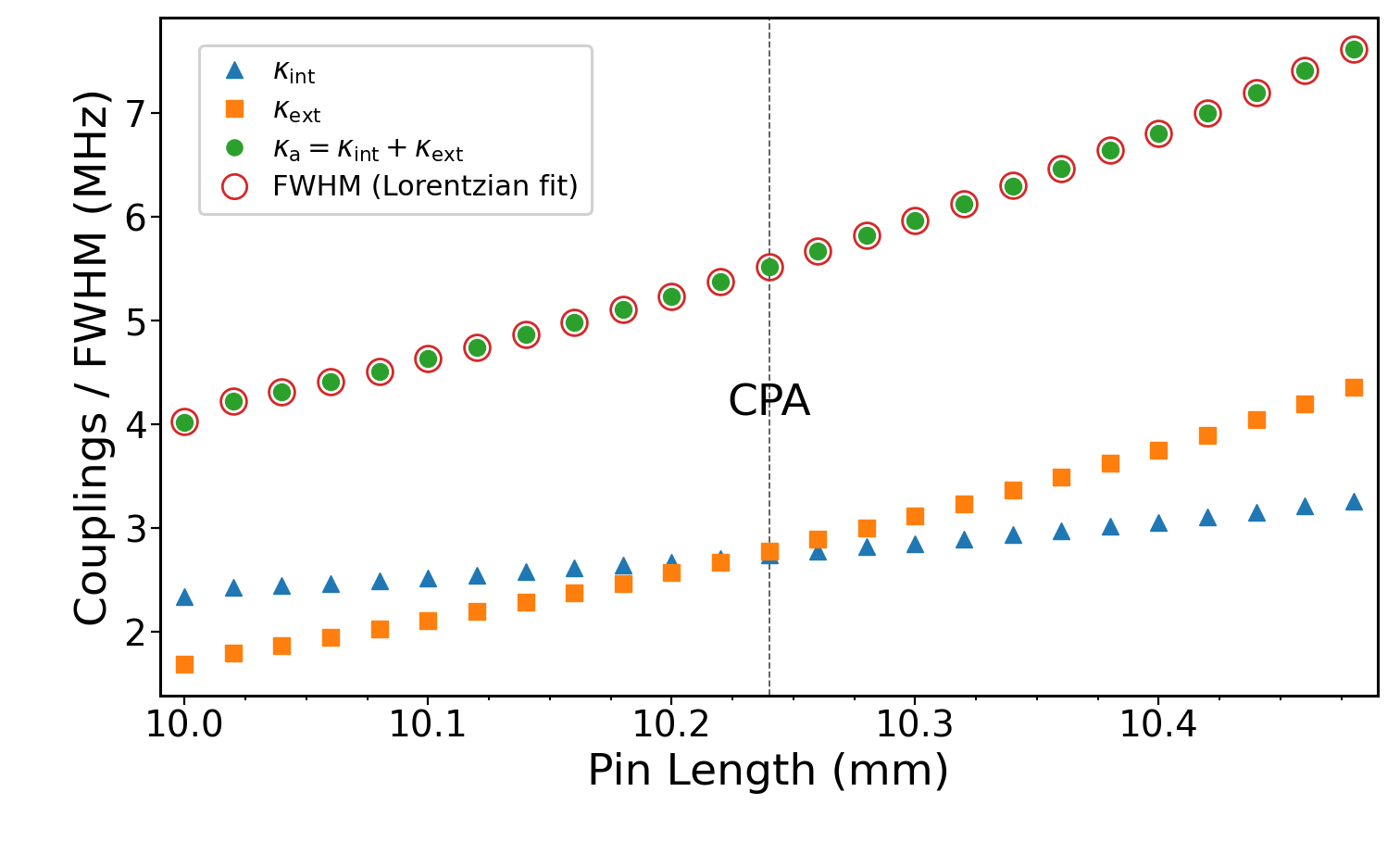}
    \caption{Fitted dissipation rates of the bare microwave cavity as a function of pin insertion length $\delta l$
    (equivalently, external dissipation rate $\kappa_{\mathrm{ext}}$). Blue: $\kappa_{\mathrm{int}}$ (nearly constant); orange: $\kappa_{\mathrm{ext}}$ (tunable); green: $\kappa_{a}=\kappa_{\mathrm{int}}+\kappa_{\mathrm{ext}}$. The rates $\kappa_{\mathrm{ext}}$, $\kappa_{\mathrm{int}}$, and $\kappa_{a}$ are extracted by fitting the spectra to Eq.~\eqref{eq:S11-power}. Red circles show the FWHM from Lorentzian fits to the measured data (equal to $\kappa_{a}$). The vertical dashed line indicates CPA (critical coupling).}

    \label{fig3}
\end{figure}

From the fits to Eq.~\eqref{eq:S11-power}, we extract $\kappa_{\mathrm{int}}$, $\kappa_{\mathrm{ext}}$, and $\kappa_a=\kappa_{\mathrm{int}}+\kappa_{\mathrm{ext}}$. Fig.~\ref{fig3} shows that $\kappa_{\mathrm{int}}$ is nearly independent of pin length (small variations attributable to changes in internal absorption with pin insertion), while $\kappa_{\mathrm{ext}}$ is strongly tunable. The FWHM of Lorentzian fits coincides with $\kappa_a$ across all couplings, including at the CPA point where $\kappa_{\mathrm{ext}}=\kappa_{\mathrm{int}}$. Taken together, these results \emph{both} verify that FWHM is set by the total decay rate $\kappa_a$ \emph{and} establish that CPA is realized precisely at critical coupling ($\kappa_{\mathrm{ext}}=\kappa_{\mathrm{int}}$), confirming that CPA modifies on-resonance amplitudes without shifting the spectral poles.

\paragraph*{Common pitfall and the pole.}
A tempting but incorrect shortcut, which is specifically used as the main idea of  Ref.~\cite{Shen2025CavityMagnon}, is to impose the CPA condition  \(a_{\mathrm{out}}=0\) and substitute \(\varepsilon_p=\sqrt{\kappa_{\mathrm{ext}}}\,a\) back into the equation of motion, Eq.~\eqref{eq:a-in-freq}, which would suggest an ``effective'' decay 
\begin{equation}
    \kappa_a' = \kappa_a-2\kappa_{\mathrm{ext}}=\kappa_{\mathrm{int}}-\kappa_{\mathrm{ext}},
\end{equation}
so one misintrepretation is that under the CPA condition where \(\kappa_{\mathrm{int}} = \kappa_{\mathrm{ext}}\),  an effective zero line width, \(\kappa_a'=0\), is achievable.  \(a_{\mathrm{out}}=0\) is valid only at the single frequency  \(\omega_p=\omega_a\) where CPA holds; it does \emph{not} change the system pole at \(\tilde{\omega}=\omega_a-i\kappa_a\) that determines the spectral width. The physical linewidth is therefore always governed by $\kappa_a = \kappa_\mathrm{{int}}+\kappa_\mathrm{{ext}}$.


\section{Reflection spectra and CPA in a cavity–magnonics system}
\label{sec:magnonics-CPA}

\subsection{Experimental data}

\begin{figure}[t]
    \centering
    \includegraphics[width=0.7\linewidth]{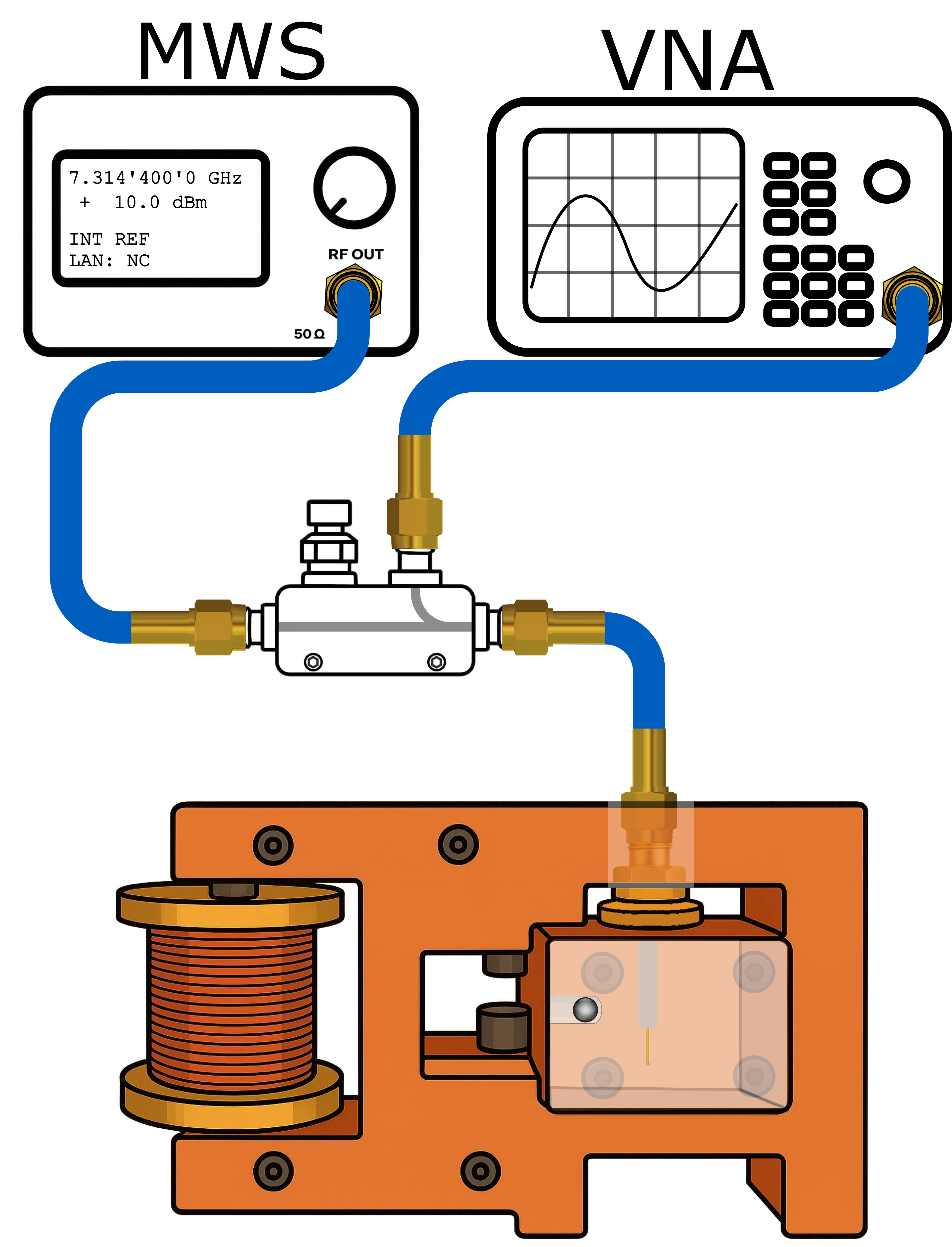}
    \caption{Schematic of the cavity–magnomechanics experiment. A weak VNA probe monitors the reflection $S_{11}$, while an independent microwave source (MWS) is power-combined onto the same feedline to resonantly drive the upper-polariton mode.}

    \label{fig4}
\end{figure}

\begin{figure}[t]
    \centering
    \includegraphics[width=\linewidth]{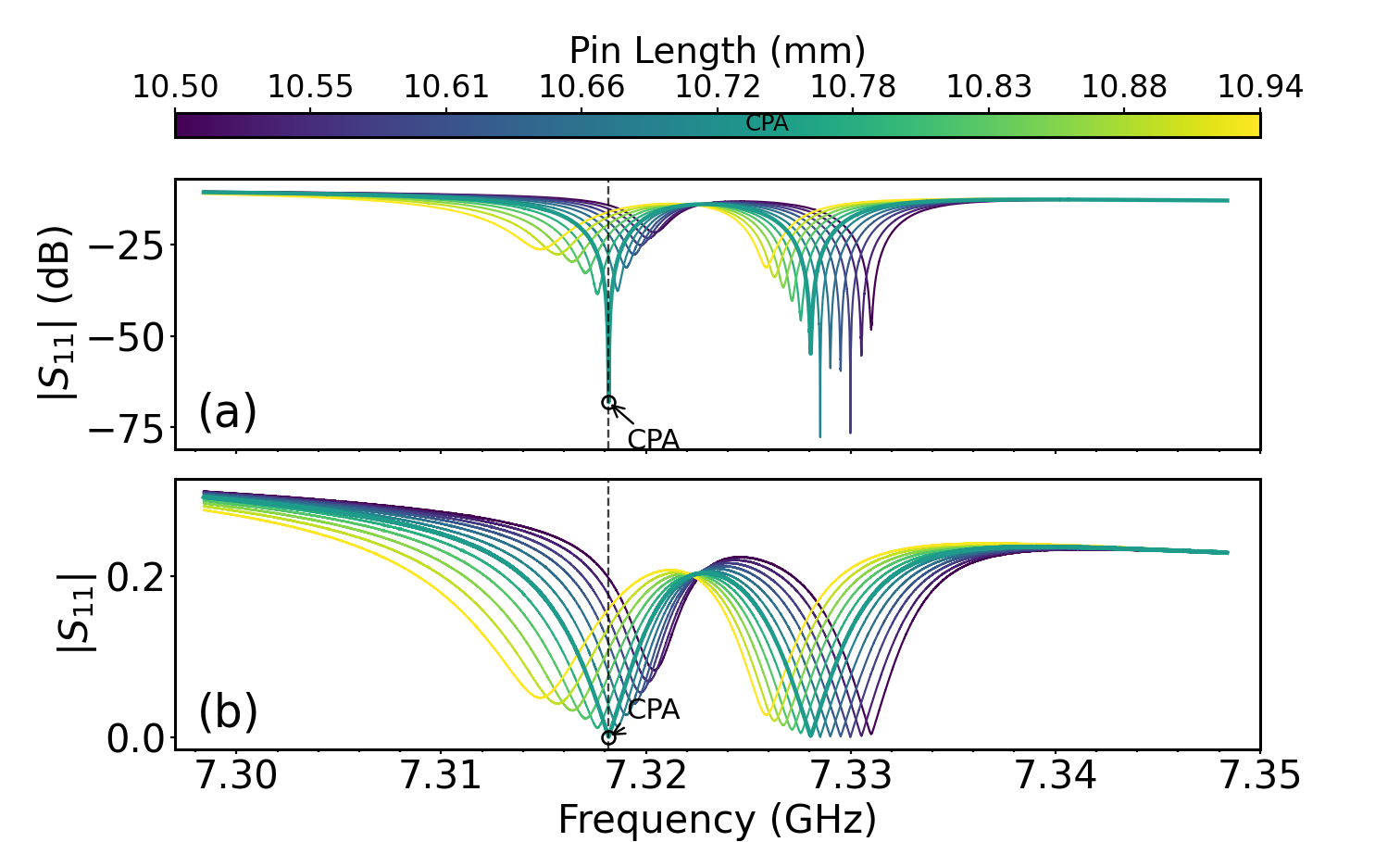}
    \caption{Reflection spectra of a single-port cavity magnonics  (Fig.~\ref{fig4}) for several pin insertion length
   $\delta l$
    (equivalently, external dissipation rate $\kappa_{\mathrm{ext}}$)
    , shown in (a) logarithmic and (b) linear units. A two-polariton modes with deepest dip in the lower mode is bolded. The vertical dashed line marks CPA of the lower polariton (critical coupling, $\kappa_{\mathrm{ext}}=\kappa_{\mathrm{int}}$), where the on-resonance dip ideally vanishes.}
    \label{fig5}
\end{figure}

We consider a single-port cavity–magnonics system with cavity resonance
$\omega_a/2\pi \approx \SI{7.32}{\giga\hertz}$.
A YIG sphere placed at the magnetic-field antinode supports magnon modes with decay rate $\kappa_{\mathrm m}/2\pi=\SI{2.49}{\mega\hertz}$.  The magnon mode couples to the cavity field via magnetic-dipole interaction, and the magnon-photon coupling strength is $g_{ma}/2\pi\approx\SI{5.50}{\mega\hertz}$.
As in the case of bare cavity, varying the insertion of the adapter pin $\delta l$ tunes the external dissipation rate $\kappa_{\mathrm{ext}}$ (Fig.~\ref{fig4}). The bias magnetic field $B_0$ is held fixed, setting the magnon frequency through
$\omega_m=\gamma B_0$ with $\gamma/2\pi=\SI{28}{\giga\hertz/\tesla}$.

Figure~\ref{fig5} shows $|S_{11}|$ measured for different $\delta l$.
On a logarithmic scale the minima on each polariton branch deepen and appear visually sharper as $\delta l$ is varied, which can suggest linewidth narrowing.
The linear-scale spectra reveal that, at specific $\delta l$ (hence $\kappa_{\mathrm{ext}}$), the on-resonance reflection approaches zero—coherent perfect absorption—\emph{without} reducing the polariton linewidths.
Polariton widths are fixed by the \emph{poles} of the hybrid susceptibility, whereas CPA concerns the \emph{zeros} of the reflection coefficient; hence CPA does not, by itself, narrow the lines.

\begin{figure}[t]
    \centering
    \includegraphics[width=\linewidth]{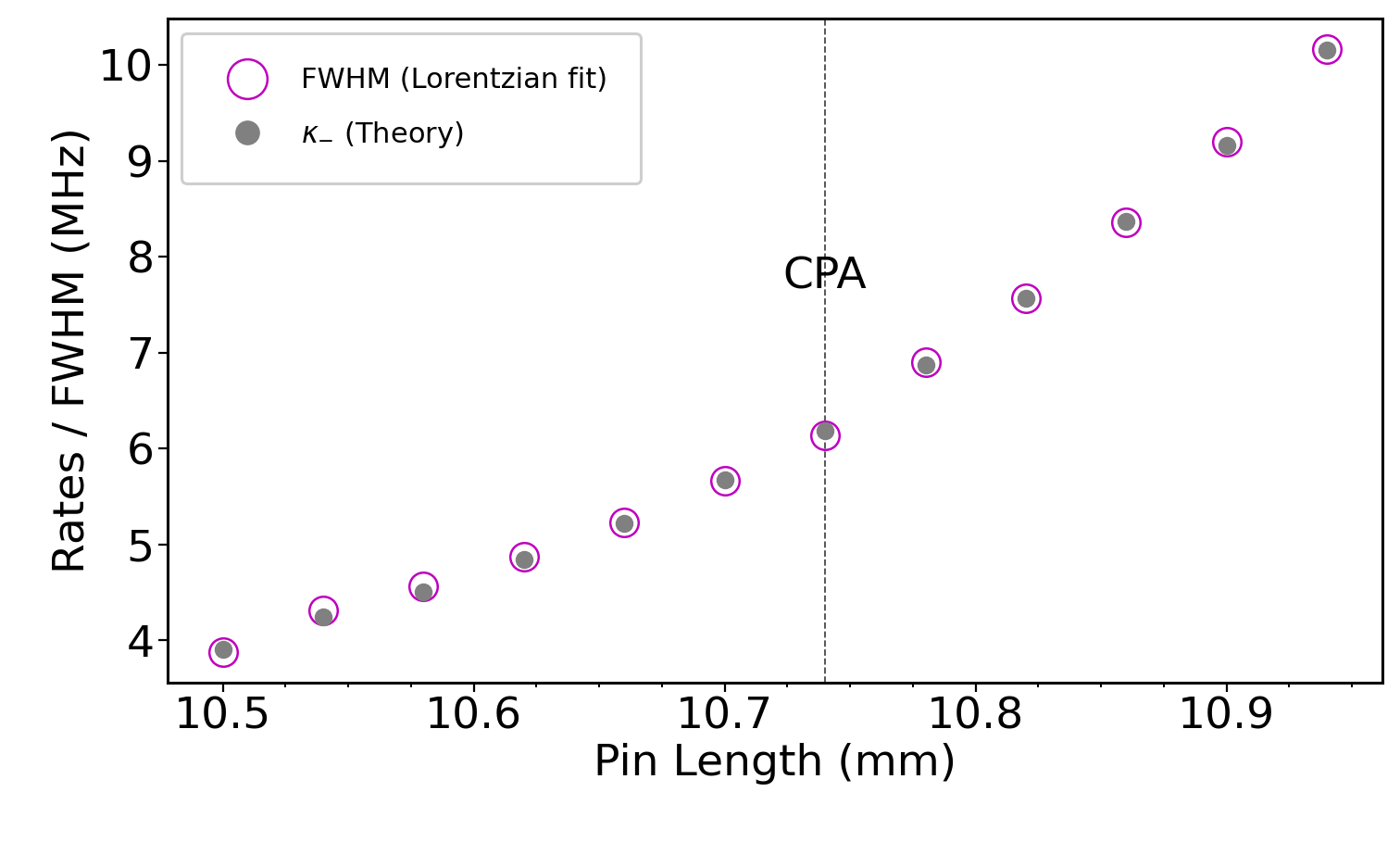}
    \caption{Lower–polariton linewidth in cavity magnonics as a function of pin insertion length $\delta l$
    (equivalently, external dissipation rate $\kappa_{\mathrm{ext}}$). FWHM from Lorentzian fits to the lower–polariton resonance in the experimental spectra (open purple circles) and theoretical prediction $\kappa_{-}=-2\,\mathrm{Im}\,\tilde{\omega}_{-}$ (filled gray circles). The vertical dashed line marks the CPA point; CPA does not reduce the linewidth.}
    \label{fig6}
\end{figure}

\subsection{Theoretical Model}

Now, we theoretically derive the reflection spectrum of a single-port cavity–magnonics system. We identify the conditions for zero reflection (CPA) and the spectral poles. Comparing these results with the analyzed experimental data extracted from Fig.~\ref{fig5}, we confirm that CPA drives the on-resonance reflection to zero without reducing the linewidth.
We consider a weak probe at frequency $\omega_p$. The Hamiltonian is
\begin{eqnarray}
\frac{H}{\hbar}
&=& \omega_a \hat a^\dagger \hat a
+ \omega_m \hat m^\dagger \hat m
+ g_{ma}\!\left(\hat a \hat m^\dagger + \hat a^\dagger \hat m\right)\nonumber\\
&&+ i\sqrt{\kappa_{\mathrm{ext}}}\,\varepsilon_p\!\left(\hat a^\dagger e^{-i\omega_p t}-\hat a e^{i\omega_p t}\right),
\label{eq:Hhyb}
\end{eqnarray}
with $\kappa_a=\kappa_{\mathrm{int}}+\kappa_{\mathrm{ext}}$ and $\kappa_m$.
Neglecting noise,
\begin{align}
\dot{\hat a} &= -\!\left(\tfrac{\kappa_a}{2}+i\omega_a\right)\hat a - i g_{ma}\hat m
               + \sqrt{\kappa_{\mathrm{ext}}}\,\varepsilon_p e^{-i\omega_p t}, \label{eq:QLEa}\\
\dot{\hat m} &= -\!\left(\tfrac{\kappa_m}{2}+i\omega_m\right)\hat m - i g_{ma}\hat a .
\label{eq:QLEm}
\end{align}
In frequency space,
\begin{equation}
a(\omega_p)=
\frac{\sqrt{\kappa_{\mathrm{ext}}}\bigl[\tfrac{\kappa_m}{2}+i(\omega_m-\omega_p)\bigr]\varepsilon_p}
{\bigl[\tfrac{\kappa_m}{2}+i(\omega_m-\omega_p)\bigr]\bigl[\tfrac{\kappa_a}{2}+i(\omega_a-\omega_p)\bigr]+g_{ma}^{2}}.
\label{eq:chi}
\end{equation}
With input-output theory, Eq.~\eqref{input-output}, the reflection spectrum is obtained as
\begin{widetext}
\begin{equation}
S_{11}(\omega_p)=
\frac{\bigl[\tfrac{\kappa_m}{2}+i(\omega_m-\omega_p)\bigr]\bigl[\tfrac{\kappa_{\mathrm{int}}-\kappa_{\mathrm{ext}}}{2}+i(\omega_a-\omega_p)\bigr]
+ g_{ma}^{2}}
{\bigl[\tfrac{\kappa_m}{2}+i(\omega_m-\omega_p)\bigr]\bigl[\tfrac{\kappa_a}{2}+i(\omega_a-\omega_p)\bigr]+g_{ma}^{2}} .
\label{eq:S11}
\end{equation}
\end{widetext}

\paragraph*{CPA at the polariton frequencies (split branch).}
For the resonant case $\omega_a=\omega_m\equiv\omega_0$, CPA corresponds to \emph{zeros of the numerator} of Eq.~\eqref{eq:S11}. Seeking split-frequency zeros at $\omega_p^{\mathrm{CPA}}=\omega_0\pm\Delta, $ with $\Delta\neq0$ and separating real/imaginary parts yields
\begin{equation}\label{eq:CPAcond}
\kappa_{\mathrm{ext}}=\kappa_{\mathrm{int}}+\kappa_m,
\end{equation}
\begin{equation}\label{eq:CPAsplit}
\Delta=\sqrt{\,g_{ma}^{2}-\frac{\kappa_m^{2}}{4}\,},
\end{equation}
which respectively can be considered as the cirtical coupling and resonance conditions of the cavity magnonics.
Thus, for $g_{ma}>\kappa_m/2$, perfect absorption occurs at the two polariton frequencies when the external coupling equals the total internal loss of the hybrid.

\paragraph*{Poles and linewidths.}
The poles (roots of the denominator of Eq.~\eqref{eq:S11}) are
\begin{eqnarray}
\tilde{\omega}_{\pm}&=&\omega_{\pm}-i\frac{\kappa_{\pm}}{2}
=\frac{\omega_a+\omega_m}{2}-\frac{i}{4}(\kappa_a+\kappa_m)\nonumber\\
&& \pm\ \frac{1}{2}\sqrt{\Big[(\omega_a-\omega_m)-\tfrac{i}{2}(\kappa_a-\kappa_m)\Big]^2+4g_{ma}^2}.\nonumber\\
\label{eq:poles}
\end{eqnarray}

The observable half-widths are $\kappa_{\pm}/2=-\mathrm{Im}\,\tilde{\omega}_{\pm}$, so the FWHM equals $\kappa_{\pm}$  and is fixed the \emph{denominator}. Satisfying Eqs.~\eqref{eq:CPAcond}–\eqref{eq:CPAsplit} (zeros of the numerator) drives $|S_{11}|$ to zero at specific frequencies but does not move the poles or narrow the polariton linewidths.

For further analysis we focus on the lower polariton (a two-polariton modes with deepest dip in the lower mode is bolded in Fig~\ref{fig5}). Figure~\ref{fig6} summarizes the FWHM extracted from Lorentzian fit to the lower polariton mode and compares it with the prediction from Eq.~\eqref{eq:poles} using parameters obtained from simultaneous fits to both branches. The agreement confirms that the linewidth is not reduced under CPA.

\begin{figure}[t]
    \centering
    \includegraphics[width=\linewidth]{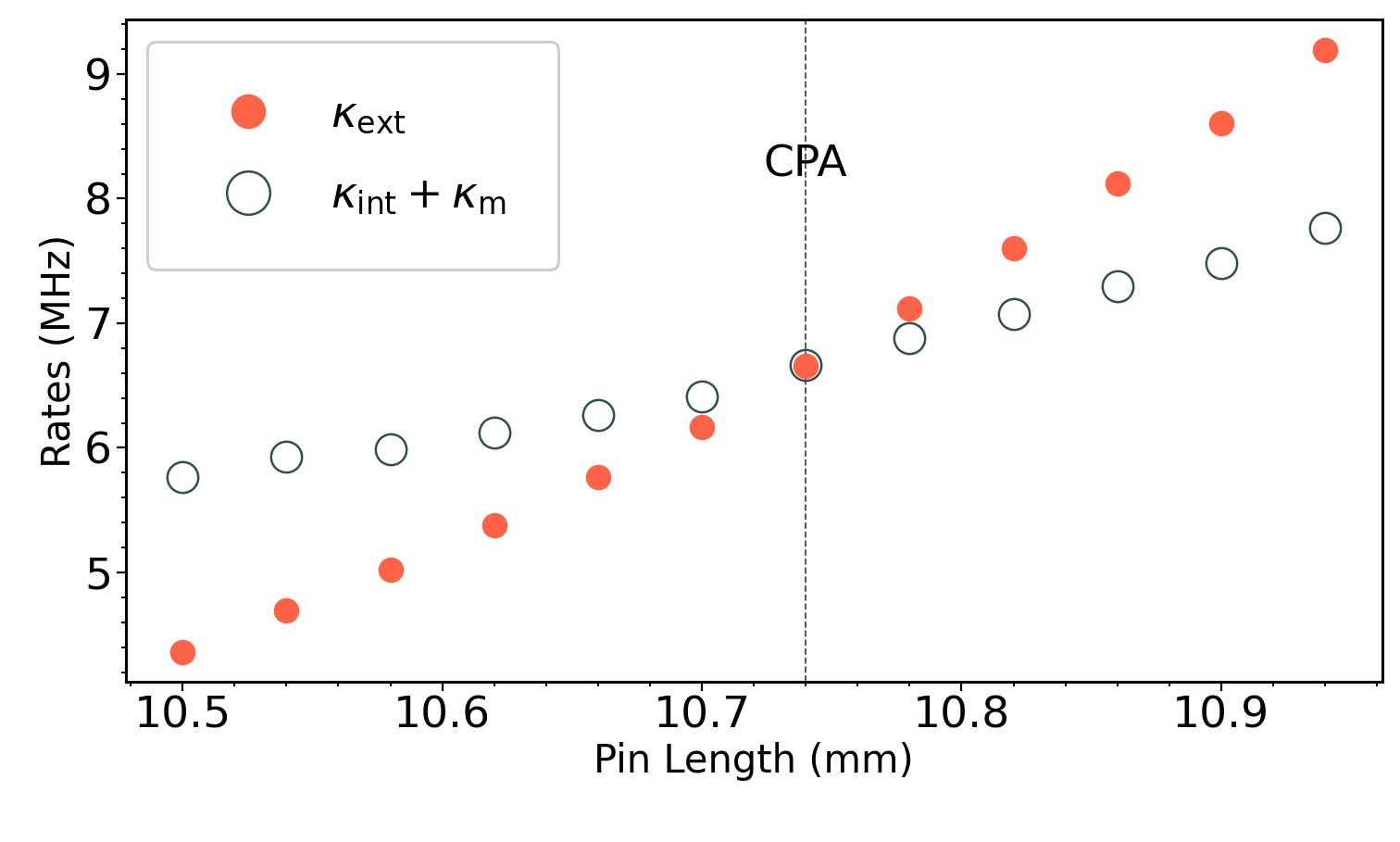}
    \caption{Verification of the first CPA condition (critical coupling) in a cavity–magnon system, $\kappa_{\mathrm{ext}}=\kappa_{\mathrm{int}}+\kappa_{m}$, as a function of pin length $\delta l$ (which tunes the external dissipation rate $\kappa_{\mathrm{ext}}$). The condition is nearly satisfied at the CPA point, $\delta l \approx \SI{10.740}{\milli\metre}$, where the on-resonance reflection approaches zero. These experimental data confirm the theoretical CPA condition, Eq.~\eqref{eq:CPAcond}.}
    \label{fig-new1}
\end{figure}

\begin{figure}[t]
    \centering
    \includegraphics[width=\linewidth]{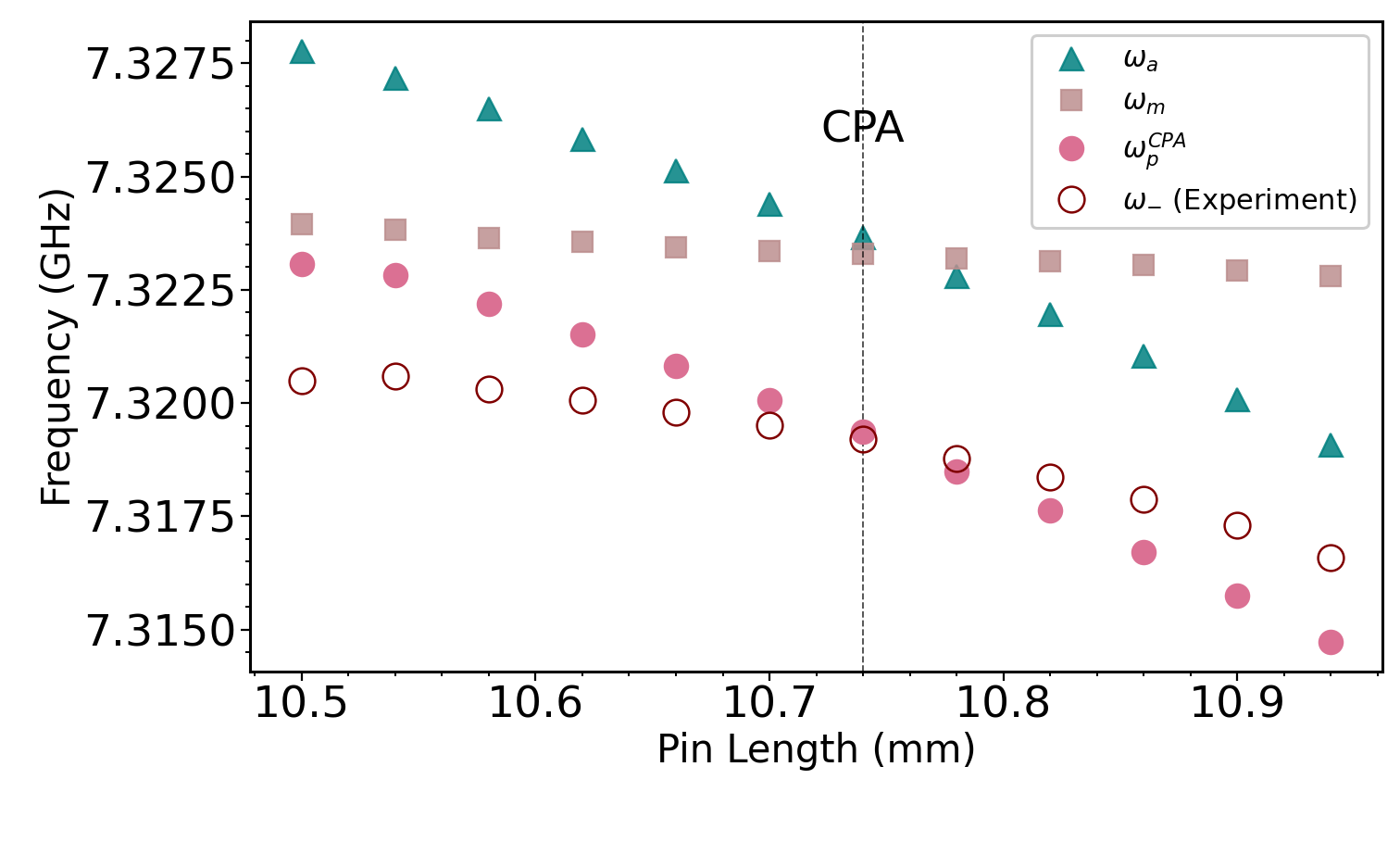}
    \caption{Verification of the second CPA condition (resonance condition) in a cavity magnonic system, $\omega_p^{\mathrm{CPA}}=\omega_a-\sqrt{g_{ma}^2-(\kappa_m/2)^2}$, versus pin insertion length $\delta l$ (which tunes the external dissipation rate $\kappa_{\mathrm{ext}}$). Markers indicate the cavity ($\omega_a$) and magnon ($\omega_m$) frequencies, extracted by fitting data to the reflection spectrum, Eq.~\eqref{eq:S11}. Filled circles denote the frequency required for zero reflection (second CPA condition) at each $\delta l$, while open circles show the measured lower-polariton frequency for each $\delta l$. The vertical dashed line marks the $\delta l$ at which CPA occurs. At resonance ($\omega_a=\omega_m$), CPA is realized when the Polariton frequency reaches the CPA-predicted  frequency $\omega_p^{\mathrm{CPA}}$.
 }
    \label{fig-new2}
\end{figure}

To experimentally check both CPA conditions, Eqs.~\eqref{eq:CPAcond}–\eqref{eq:CPAsplit},   Fig.~\ref{fig-new1} shows the rates extracted from fits: at the CPA point the critical-coupling relation $\kappa_{\mathrm{ext}}=\kappa_{\mathrm{int}}+\kappa_m$ [Eq.~\eqref{eq:CPAcond}] is satisfied. Complementarily, Fig.~\ref{fig-new2} tests the frequency condition: for $\omega_a\simeq\omega_m\equiv\omega_0$, the CPA zeros occur at $\omega_p^{\mathrm{CPA}}=\omega_0\pm\Delta$ with $\Delta=\sqrt{g_{ma}^2-(\kappa_m/2)^2}$ [Eq.~\eqref{eq:CPAsplit}]; the measured lower-polariton frequency follows this prediction at the observed CPA point.

\section{Cavity magnomechanics under the CPA condition}

In this section we investigate magnomechanical interactions in the presence of CPA. By driving either the cavity or the magnon mode one excites a polariton; the resulting magnetostrictive response mediates a magnon–phonon interaction. In the dispersive regime this interaction is directly analogous to the radiation-pressure coupling in cavity optomechanics and gives rise to magnomechanically induced transparency (MMIT) \cite{Zhang2016, Potts2021}. In our experiment we excite the polariton by driving the microwave cavity with a strong microwave source (Fig.~\ref{fig4}).

\begin{figure}[t]
    \centering
    \includegraphics[width=0.8\linewidth]{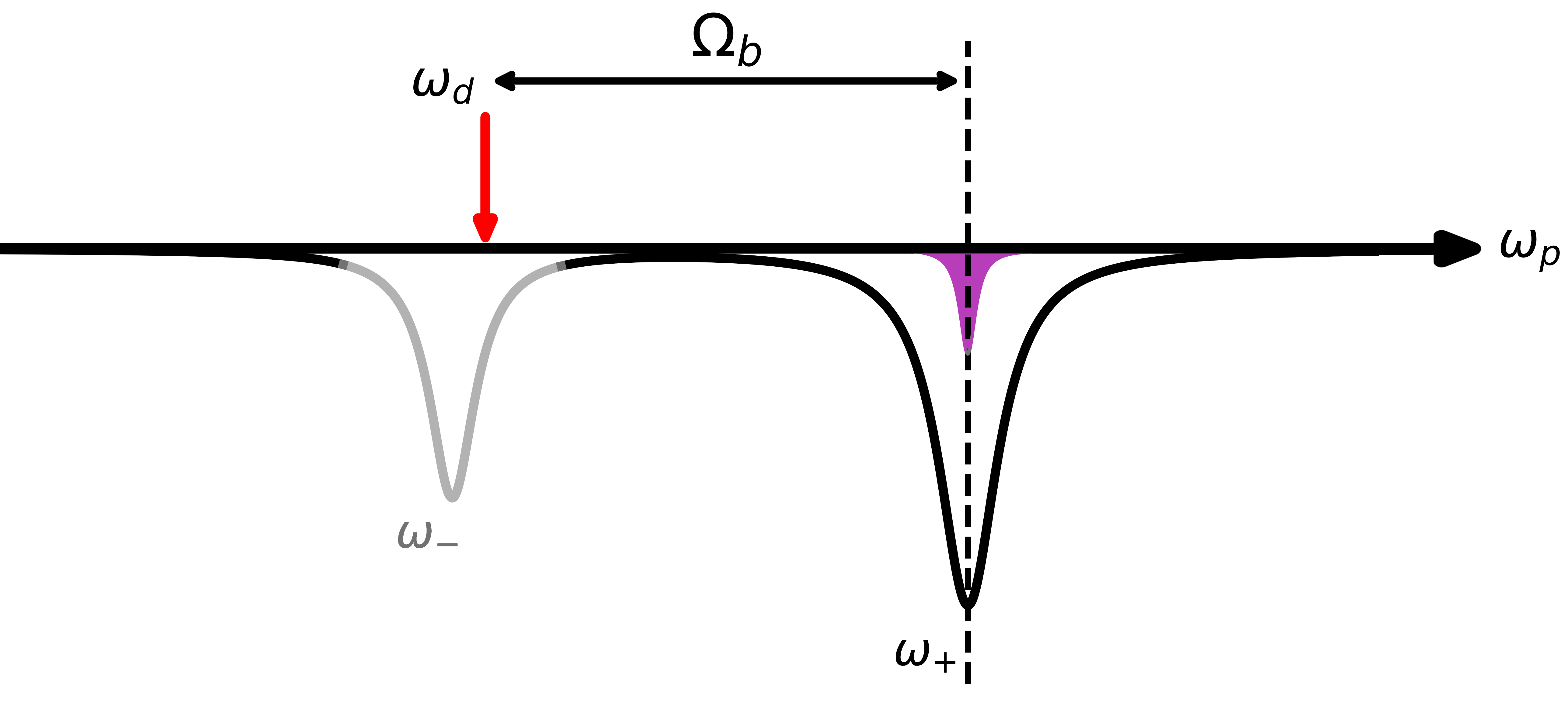}
   \caption{Frequency schematic based on the measurement. The red arrow marks the microwave source (MWS) drive at $\omega_d$, which excites the upper polariton mode at $\omega_{+}$ (dashed line). The lower polariton $\omega_{-}$ is shown in gray. $\Omega_b$ denotes the mechanical frequency, and $\omega_p$ is the probe frequency axis.}

    \label{last}
\end{figure}

\begin{figure*}[t]
  \centering
  \includegraphics[width=\textwidth]{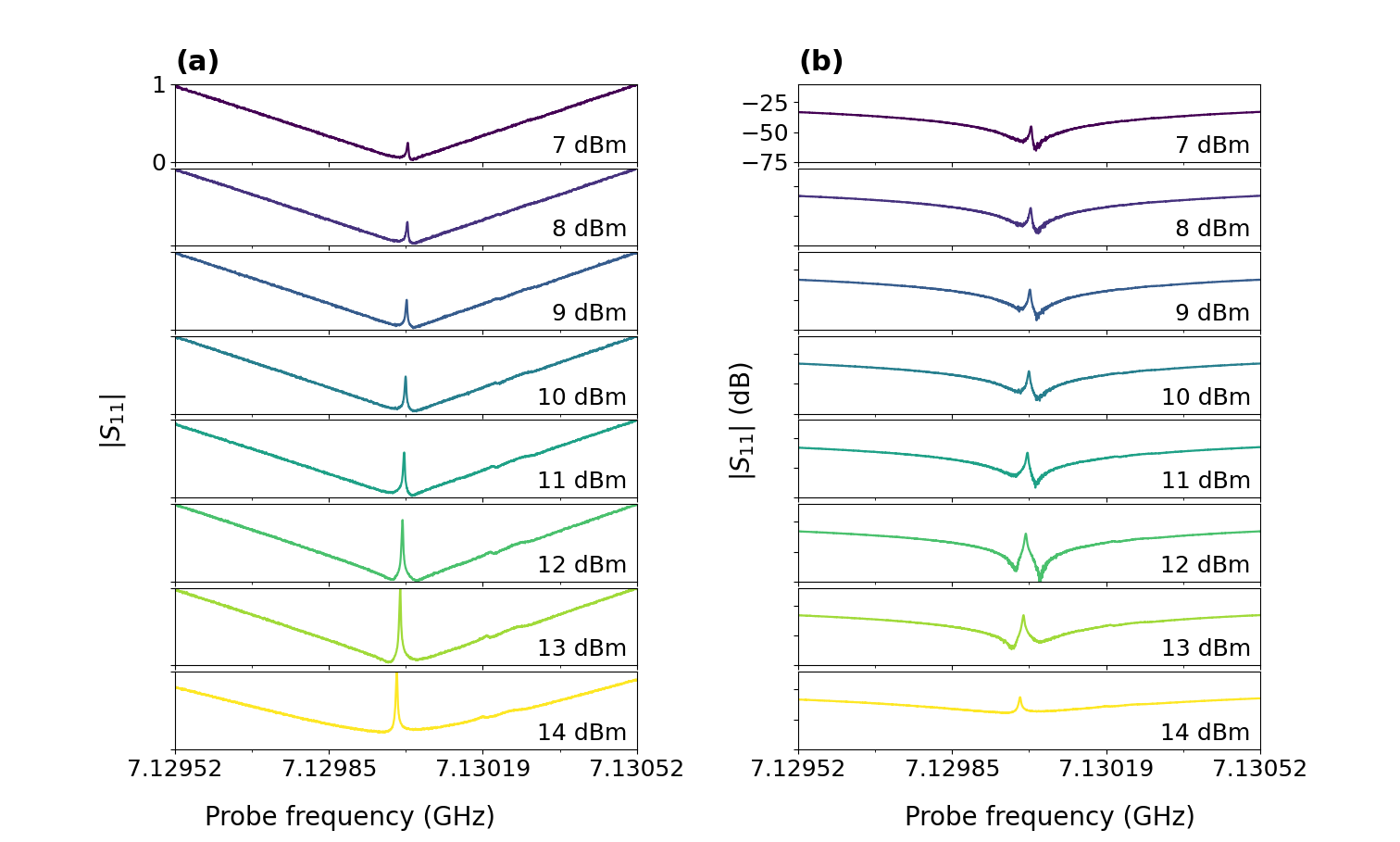}
  \caption{Reflection spectra $|S_{11}|$ of the upper polariton versus probe frequency for drive powers 7–14~dBm, shown over a 1~MHz span. The cavity is driven with a strong tone red-detuned by $\Omega_d/2\pi = 15.6$~MHz (low-frequency/left side) from the upper polariton, so MMIT appears on its red shoulder while the polariton is tuned to CPA. Because changing the drive power shifts the polariton frequency, we retune the bias magnetic field $B_0$ at each power to keep the mechanical mode at the CPA point (i.e., maintain $\Delta_{-}=\Omega_b$). \textbf{(a)} linear scale. \textbf{(b)} logarithmic. The near-zero minimum at CPA can mimic a doublet in logarithmic, but the linear-scale data show a single mode with a transparency window, not a genuine splitting.}
  \label{fig-mec1}
\end{figure*}

\begin{figure*}[t]
  \centering
  \includegraphics[width=\textwidth]{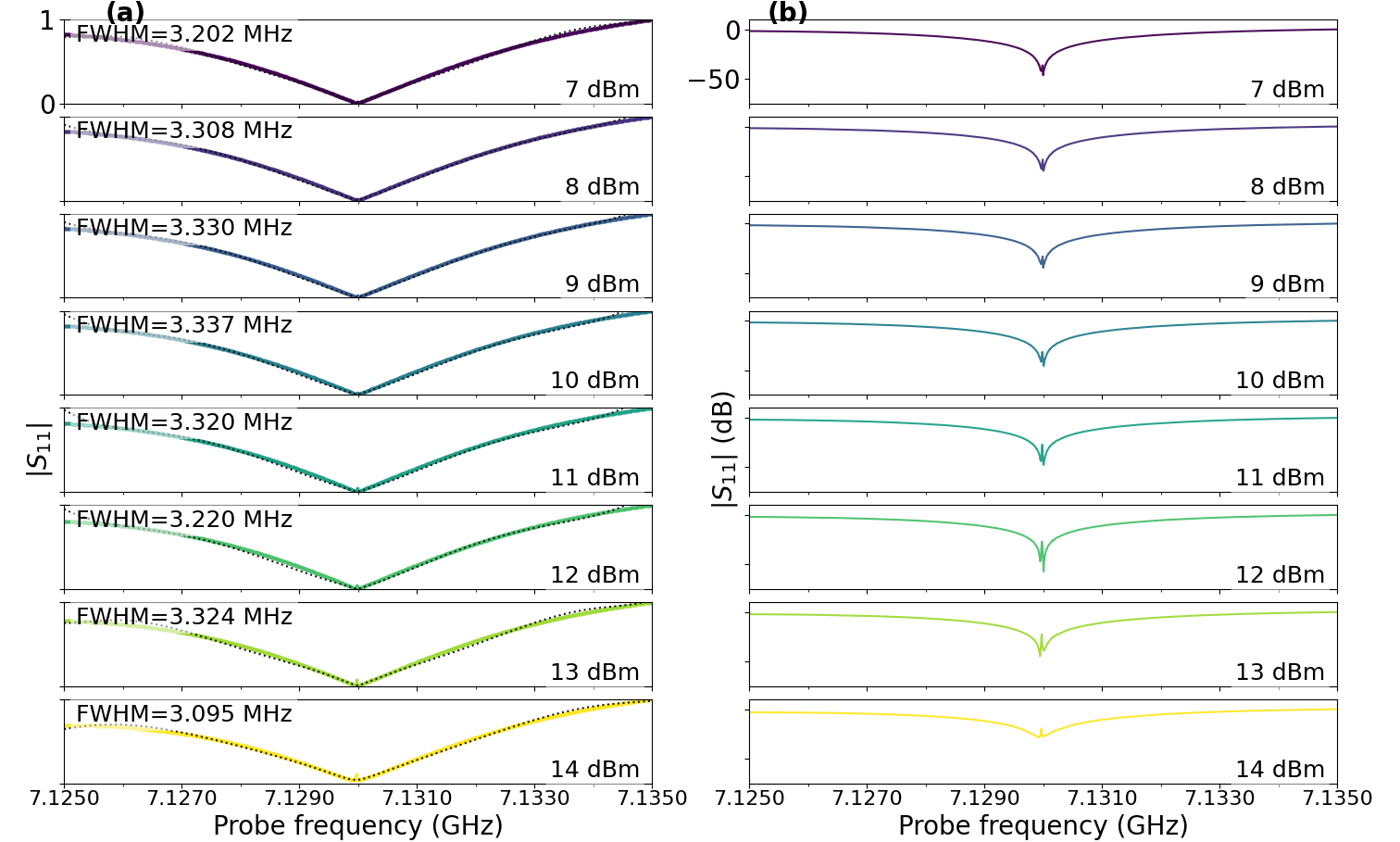}
\caption{Reflection spectra $|S_{11}|$ of the upper polariton versus probe frequency for drive powers 7–14~dBm, shown over a 10~MHz span. The cavity is driven with a strong tone red–detuned from the upper polariton by $\Delta_{-}/2\pi=\Omega_{b}/2\pi=15.6~\text{MHz}$, so MMIT appears on its red shoulder while the polariton is tuned to CPA. Because the drive power shifts the polariton frequency, the bias field $B_{0}$ is retuned at each power to maintain the CPA operating point. \textbf{(a)} Linear magnitude; dotted curves are fits, and the extracted FWHM for each trace is annotated (all $\approx 3.3$~MHz). \textbf{(b)} Logarithmic magnitude. The wider 10~MHz window, which captures the full width of the upper polariton, shows a single resonance with a weak transparency feature (more apparent in \textbf{(b)})—no genuine normal-mode splitting. CPA deepens the dip but does not narrow the linewidth; the FWHM remains essentially constant across drive powers.}
\label{fig-mec2}
\end{figure*}

We operate with the drive \emph{red-detuned} from the upper polariton by $\Delta_{-}=\Omega_b=\SI{15.6}{\mega\hertz}$, where we define
$\Delta_{-}\equiv \omega_{-}-\omega_d$ (with $\omega_d$ the drive frequency and $\omega_{-}$ the lower–polariton frequency) and
$\Omega_b$ as the mechanical-mode frequency. This places the probe on the low-frequency side of the upper polariton, so that MMIT appears on its red shoulder (Fig.~\ref{last}). Throughout, the polariton is tuned to satisfy the CPA condition. Representative \emph{reflection} ($S_{11}$) spectra of the upper polariton, in both linear and logarithmic scales, are shown in Figs.~\ref{fig-mec1} and \ref{fig-mec2} for drive powers from \SI{7}{dBm} to \SI{12}{dBm}. The two panels use different frequency spans—\SI{1}{\mega\hertz} in Fig.~\ref{fig-mec1} and \SI{10}{\mega\hertz} in Fig.~\ref{fig-mec2}. As expected under CPA, the on-resonance reflection at the bottom of the polariton approaches zero. While logarithmic plots over a narrow span can give the impression of a splitting (as incorrectly concluded in Ref.~\cite{Shen2025CavityMagnon} and as can be seen at 12 dBm drive in Fig.~\ref{fig-mec1}b, based solely on logarithmic representations), the \emph{linear}-scale, Fig.~\ref{fig-mec1}a, spectra clearly exhibit MMIT \emph{without} any genuine splitting. CPA deepens the polariton dip; under drive, the resulting sharp minimum can resemble a doublet on a logarithmic scale. However, the wider-span linear response (Fig.~\ref{fig-mec2}), which captures the full width of the upper-polariton mode, clearly indicates a single mode. If a genuine splitting were present, it would also appear in this figure.

\begin{figure*}[t]
  \centering
  \includegraphics[width=\textwidth]{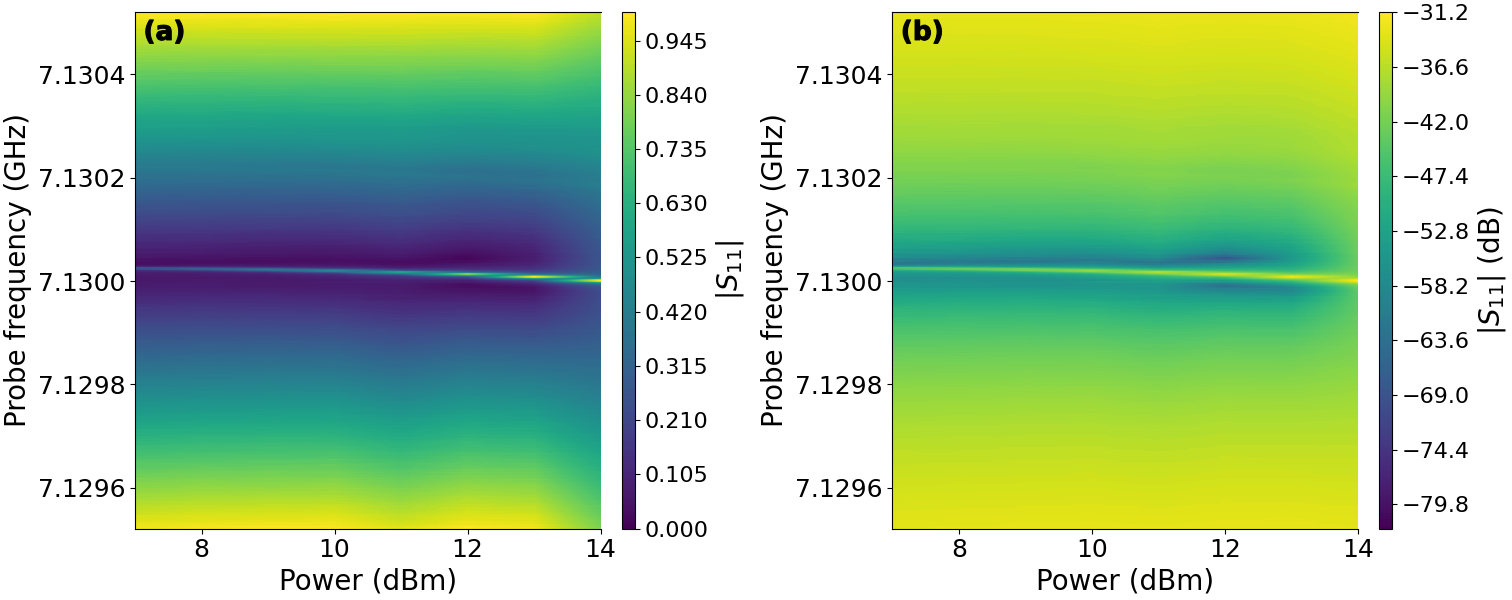}
\caption{Contour plots of the upper–polariton reflection \(|S_{11}|\) versus probe frequency and drive power (\SI{7}{–}\SI{14}{dBm}). The cavity is driven red-detuned by \(\Omega_d/2\pi=\SI{15.6}{MHz}\) (low-frequency side) with the polariton tuned to CPA. \textbf{(a)} linear magnitude. \textbf{(b)} logarithmic. The narrow ridge marks MMIT at CPA; no genuine normal-mode splitting (the logarithmic “doublet’’ is a plotting artifact).}
  \label{fig-mec3}
\end{figure*}

Figure~\ref{fig-mec3} presents a contour plot of $S_{11}$ versus drive power and probe frequency. Again, logarithmic views may suggest an apparent splitting near CPA, whereas the corresponding linear-scale data show MMIT without any sign of mode splitting. Thus, the interpretation in Ref.~\cite{Shen2025CavityMagnon} is \emph{incorrect}: CPA does not cause linewidth reduction leading to true normal-mode splitting; the apparent features arise from plotting logarithmically and from the sharpened CPA dip.

\begin{figure*}[t]
  \centering
  \includegraphics[width=\textwidth]{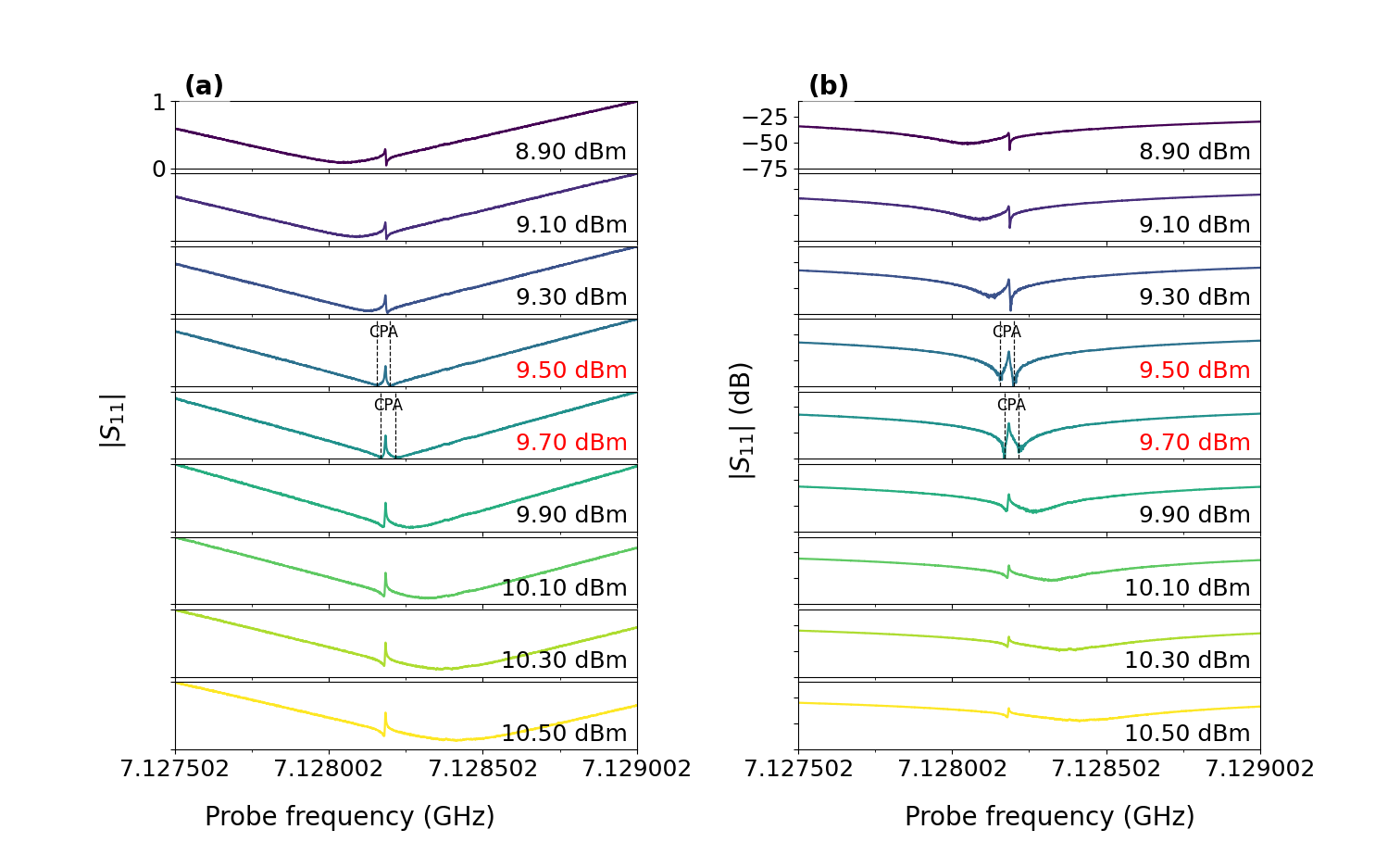}
    \caption{Reflection spectra \(|S_{11}|\) of the upper polariton versus probe frequency for drive powers \SI{8.90}{dBm}–\SI{10.50}{dBm}. With the bias field fixed, increasing power shifts the polariton; CPA is reached only near \SI{9.5}{dBm} and \SI{9.7}{dBm}, where the dip sharpens. \textbf{(a)} linear magnitude. \textbf{(b)} logarithmic. Near CPA the logarithmic traces can resemble a doublet, but the linear-scale spectra show a single resonance with an MMIT feature—no true normal-mode splitting.}

  \label{fig-mec4}
\end{figure*}

We further measured the response while increasing the drive power \emph{without} adjusting the bias magnetic field. In this case, power-induced shifts of the magnon frequency move the polariton due to magnon Kerr nonlinearity \cite{Zhang2016, Shen2022}, so that CPA is not maintained at all drive powers. The logarithmic spectra in Fig.~\ref{fig-mec4} can resemble polaromechanical normal-mode splitting (as claimed in Ref.~\cite{Shen2025CavityMagnon}), yet the linear-scale traces do not exhibit true splitting. The apparent “splitting’’ near \SI{9.5}{dBm}–\SI{9.7}{dBm} coincides with sharper dips when CPA is satisfied.

\begin{figure*}[t]
  \centering
  \includegraphics[width=\textwidth]{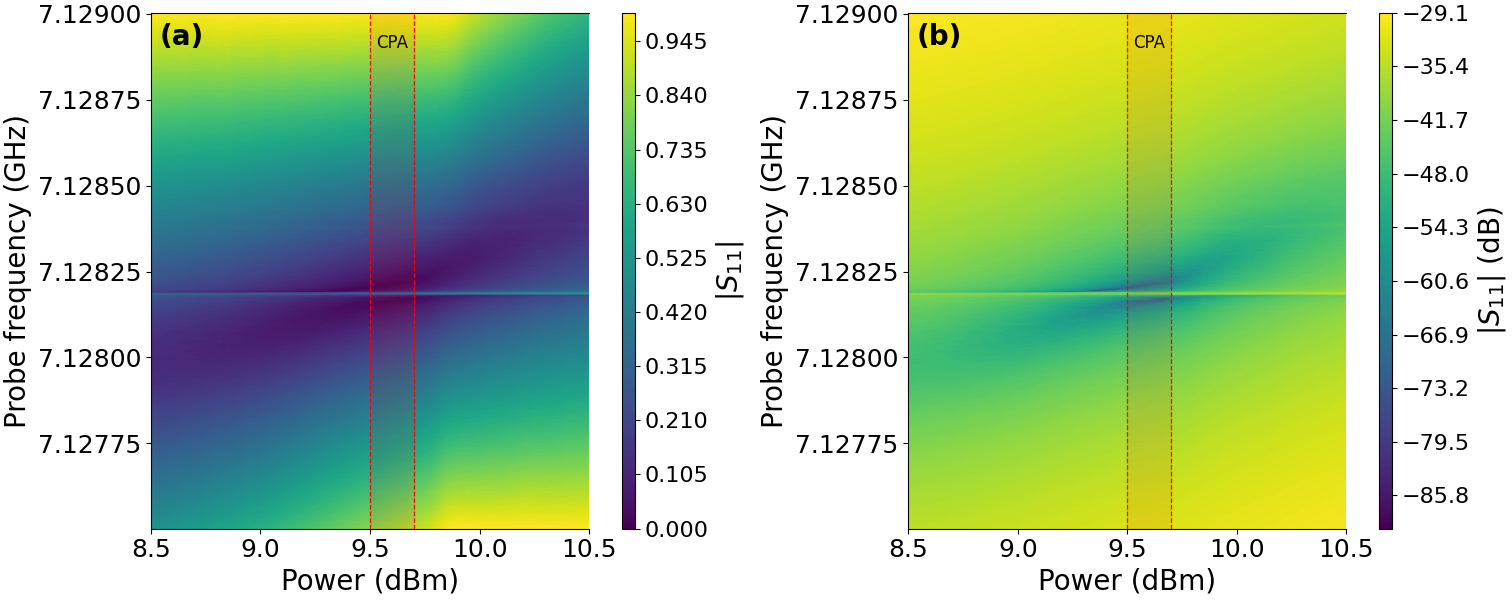}
    \caption{Contour plots of the upper–polariton reflection $|S_{11}|$ versus probe frequency and drive power with the bias field fixed.
    \textbf{(a)} linear magnitude.
    \textbf{(b)} logarithmic.
    The shaded region between the red dashed lines indicates near-zero reflection, corresponding to the onset of coherent perfect absorption (CPA), reached only near \SIrange{9.5}{9.7}{dBm}.}
  \label{fig-mec5}
\end{figure*}

Finally, Fig.~\ref{fig-mec5} shows a contour plot of the reflection spectra versus drive power and probe frequency for the same protocol (increasing drive power without retuning the magnon). As in Fig.~\ref{fig-mec4}, logarithmic plots can misleadingly suggest splitting at CPA, but the linear-scale data reveal no genuine normal-mode doublet. Together with the discussion above on CPA, these observations emphasize that CPA does not imply linewidth reduction, and that the conclusion of polaromechanical normal-mode splitting drawn in Ref.~\cite{Shen2025CavityMagnon} is not acceptable: claims of splitting require corroboration from both linear- and logarithmic spectra over a sufficiently wide frequency range. Here, the linear-scale results in Fig.~\ref{fig-mec5} show no evidence of true splitting.

\section{Conclusion}
We have revisited how coherent perfect absorption (CPA) affects spectral widths in single-port cavities and cavity–magnon hybrids, combining input–output theory with experiments displayed in both linear and logarithmic units. The key outcome is that CPA creates a \emph{zero} of the reflection amplitude without shifting the \emph{poles} that set the linewidth: the FWHM is governed by the total decay rate $\kappa_a=\kappa_{\mathrm{int}}+\kappa_{\mathrm{ext}}$ (or, in the hybrid case, by the polariton poles) and is \emph{not} reduced at CPA. Visual “sharpening’’ seen in logarithmic plots over narrow spans reflects amplitude zeros, not linewidth narrowing.

We further studied \emph{magnomechanical} interactions under the CPA condition. In this regime, logarithmic spectra can display sharp dips that mimic a doublet; however, linear-scale data over a wider span reveal a single polariton with a transparency feature (MMIT), not a genuine polaromechanical normal-mode splitting. Accordingly, the splitting reported in Ref.~\cite{Shen2025CavityMagnon} is a misinterpretation arising from logarithmic sharp minima at CPA rather than from true pole splitting.

Practically, linewidths and mode structure should be inferred from linear-scale fits or from the pole locations of the response, with logarithmic plots used for dynamic range only when accompanied by linear-scale spectra. These results correct the misconception that CPA provides a route to linewidth suppression or polaromechanical normal-mode splitting in the linear, weak-probe regime, and establish robust guidelines for analysis and presentation in cavity magnonics and related platforms.

\clearpage          

\bibliography{apssamp}

\end{document}